\documentclass[useAMS,usenatbib]{mn2e} 
\usepackage{aas_macros}
\usepackage{graphicx}
\usepackage{deluxetable}

\title{SgrA* emission at 7mm: variability and periodicity}
\author[P.P.B.Beaklini and Z.Abraham]
{Pedro Paulo B. Beaklini$^{1}$%{E-mail:
\newauthor
Zulema Abraham$^1$\\
$^1$Instituto de Astronomia,
Geof\'{\i}sica e Ci\^{e}ncias Atmosf\'{e}ricas, Universidade de S\~{a}o Paulo.\\
Rua do Mat\~{a}o 1226, 05508-090, S\~{a}o Paulo/SP, Brazil.}%\\
\date{Released 2011 Xxxxx XX}
\begin{document}

\maketitle

\begin{abstract}
We present the result of 6 years monitoring of SgrA*, radio source associated to the supermassive black hole at the centre of the Milky Way. Single dish observations were performed with the Itapetinga radio telescope at 7 mm, and the contribution of the SgrA complex that surrounds SgrA* was subtracted and used as instantaneous calibrator.  The observations  were alternated every 10 min with those of the HII region SrgB2, which was also used as a calibrator. The reliability of the detections was tested  comparing them with simultaneous observations using interferometric techniques. During the observing period we detected a continuous increase in the SgrA* flux density starting in 2008, as well as  variability in  timescales of days and strong intraday fluctuations. We investigated if the continuous increase in flux density is compatible with free-free emission from the tail of the disrupted compact cloud that is falling towards SgrA* and concluded that the increase is most probably intrinsic to SgrA*. Statistical analysis of the light curve using  Stellingwerf and Structure Function methods revealed the existence of two minima, $156 \pm 10$  and $220 \pm 10$ days. The same statistical tests applied to a simulated light curve constructed from two quadratic sinusoidal functions superimposed to random variability  reproduced very well the results obtained with the real light curve, if the periods were 57 and 156 days.  Moreover, when a daily sampling was used in the simulated light curve, it was possible to reproduce the 2.3 GHz structure function obtained by Falcke in 1999, which revealed the 57 days period, while the 106 periodicity found by Zhao et al in 2001 could be a resonance of this period.

\end{abstract}

\begin{keywords}
Radio continuum: general --- Galaxy: centre
\end{keywords}

\section{Introduction}

SgrA* is a compact radio source, with radius smaller than 1 AU, which is associated to the $4\times 10^6 M_{\odot}$ supermassive black hole localized  in the centre of the Milky Way \citep{eck96,men97,sch02,eis03,gil09}. Its mass and distance make the event horizon of SgrA* the largest angular size of all known black hole candidates. It is a sub-Eddington source, with bolometric luminosity $10^{-10}$ times its Eddington luminosity (see \citet{mel01}, for a review). Recently, \citet{gil12} detected a gas cloud that is approaching SgrA* in a highly eccentric Keplerian orbit. The mass of the cloud has been estimated as three earth masses and will reach the black hole in the middle of 2013, during pericentre passage. The cloud  is being elongated along the orbital path as a  consequence of tidal disruption and will finally disintegrate into small clumps. 

SgrA* has been observed since the discovery of the radio source by \citet{bal74} and its emission presents variability at all wavelengths, with time scales that can be just a few hours in the infrared \citep{gen03,ghe04,mey08,do09} and  X-rays \citep{bag01,bag03}, or days and months at radio frequencies \citep{zha92,dus94,her04}. Intra day variability (IDV) have been detected at radio wavelengths: 0.35, 0.7 and 1.3 cm \citep{yus06,li09}.

The first report of periodic behaviour in SgrA* radio emission was made by \citet{fal99}, who found a 57 day periodicity  analysing  the structure function of the 2.3 GHz light curve  obtained with the Green Bank Interferometer (GBI). At that time SgrA* was also monitored at 8.3 GHz, but no periodicity was found at this frequency. At smaller wavelengths, analyses of  Very Large Array (VLA) observations at 3.6, 2.0 and 1.3 cm lasting 20 years, detected a 106 day cycle in the SgrA* emission \citep{zha01}.  In this work the period was determined using the Fourier Method and the mean profile of the 106 day cycle was obtained after 13 cycles. The periodic behaviour was more evident at 1.3 cm,  and there was an increase of  variability amplitude with  frequency. The 57 and 106 days periods are resonant but after the first detection,  no other observation was able to  confirm them,  at these or other radio frequencies. \citet{her04}  also used the VLA to study the long term radio variability of SgrA* at 2.0, 1.3 and 0.7 cm on a 3.3 year campaign. They confirmed the increase of  variability amplitude with  frequency, but the observations did not have sufficient temporal coverage to detect periodicity. 

Several models have been invoked to explain the spectral energy distribution of SgrA*, as the  quasi-Bondi-Hoyle accretion \citep{mel92,mel94,cok99,mos06,cua06}, disks \citep{nar98,yua03} and  jets \citep{fal00,mar07}. Two models must be mentioned that try to explain the short timescale variability: the Plasmon model \citep{van66,yus06,yus08} and the Hot Spot model \citep{bro05,bro06}, while the longer timescale variability can be modelled using the jet and/or disk models. As only one model was not sufficient to explain all the observational characteristics of the SgrA* emission,  symbiotic models were proposed, as the jet-disk models \citep{yua02,kun04,kun07,jol08} and the halo model \citep{pre05}. \citet{mac06} even proposed electron scattering as an extrinsic origin to explain part of the variability at radio wavelengths.  

In an attempt to understand the emission of the innermost accretion flow, recent VLBI observations at 1.3 mm inferred  $37 \mu\rm as $ for the size of the black hole shadow \citep{doe08}. VLBI observations at shorter wavelengths do not suffer the influence of interstellar scattering and could be used to detected the event horizon and the structure of its surrounding region \citep{fal00b}. However, the emission mechanisms and the variability at centimetre and millimetre wavelengths  will only be understood through monitoring programs. 

In this paper we present single dish monitoring of the 7 mm emission of SgrA* during the last six years. In Section \ref{Observations} we describe the observations and the method of data reduction, in Section \ref{Results} we show the light curve  and discuss  variability and periodicity . Finally we present our conclusion in Section \ref{conclusion}.  

\section{Observations}
\label{Observations}

The 7 mm observations were performed between 2006 and 2012  with the $13.7$ m radome enclosed Itapetinga radio telescope, localized in S\~{a}o Paulo, Brazil. At this wavelength, the antenna half power beam width (HPBW) is $2.2$ arc min  and the radome transmission is $0.68$. The receiver, a room temperature K-band mixer, has a 1-GHz double side band, with a $\sim 700$ K of noise temperature. The calibration was made with a  noise source of known temperature and a room temperature load, which take into account the atmospheric and radome attenuation \citep{abr92}. Virgo A, which has a flux density of 11 Jy at 7 mm,  was used as the primary flux calibrator.

\begin{figure}
  \includegraphics[width=\columnwidth]{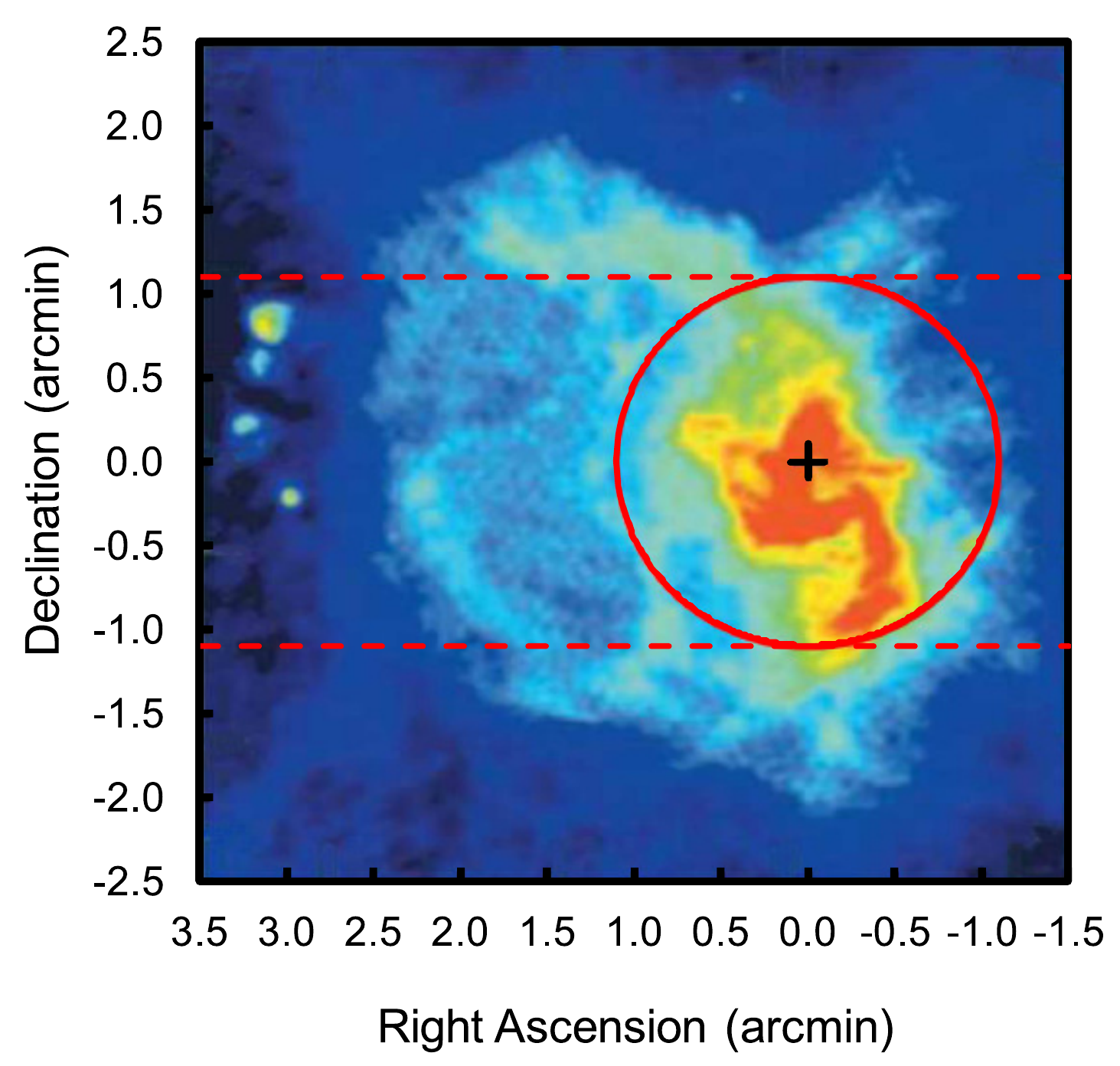}
  \caption{Map of the Galactic centre ($5^\prime \times 5^\prime$) at 6 cm \citep{yus00}, indicating  the the beam size of the Itapetinga radio telescope (red circle) and the scan direction (pointed red lines).  The position of SgrA* is indicate by a black cross.}
  \label{GC}
\end{figure}
	
The method of observation was scans in right ascension, with 30 arcmin amplitude and 20 s duration.  Only 24 arcmin were effectively utilized, the rest was lost during the inversion of the scan direction. During  each scan 81 equally spaced  data values, with integration time of $1/6$ s, were obtained. Each observation consisted of the average of 30 scans, preceded by a calibration; during an observing day, between 7 and 14 observations were obtained and averaged, always when the source was in elevation above $30^{\circ}$ and below $75^{\circ}$. The observations of SgrA* were alternated with similar  observation of the HII region SgrB2, which was used as a quasi-simultaneous secondary calibrator.
	
The main problem in the detection of radio variability from SgrA* with  single dish observation is the contribution of the continuum emission from the region surrounding this point source. As we show in  Fig. \ref{GC}, one scan with 24 arcmin amplitude and 2.2 arc min HPBW, detects not only SgrA* but  part of the continuum complex SgrA. Since only SgrA* emission is variable,  we  used SgrA as an instantaneous calibrator,  and as a result increased the variability detection capacity of our observations. To check the accuracy of this calibration and obtain a realistic value of the errors involved in the flux determination, we used a second calibrator: the strong HII region SgrB2 Main.

\begin{figure}
  \includegraphics[width=\columnwidth]{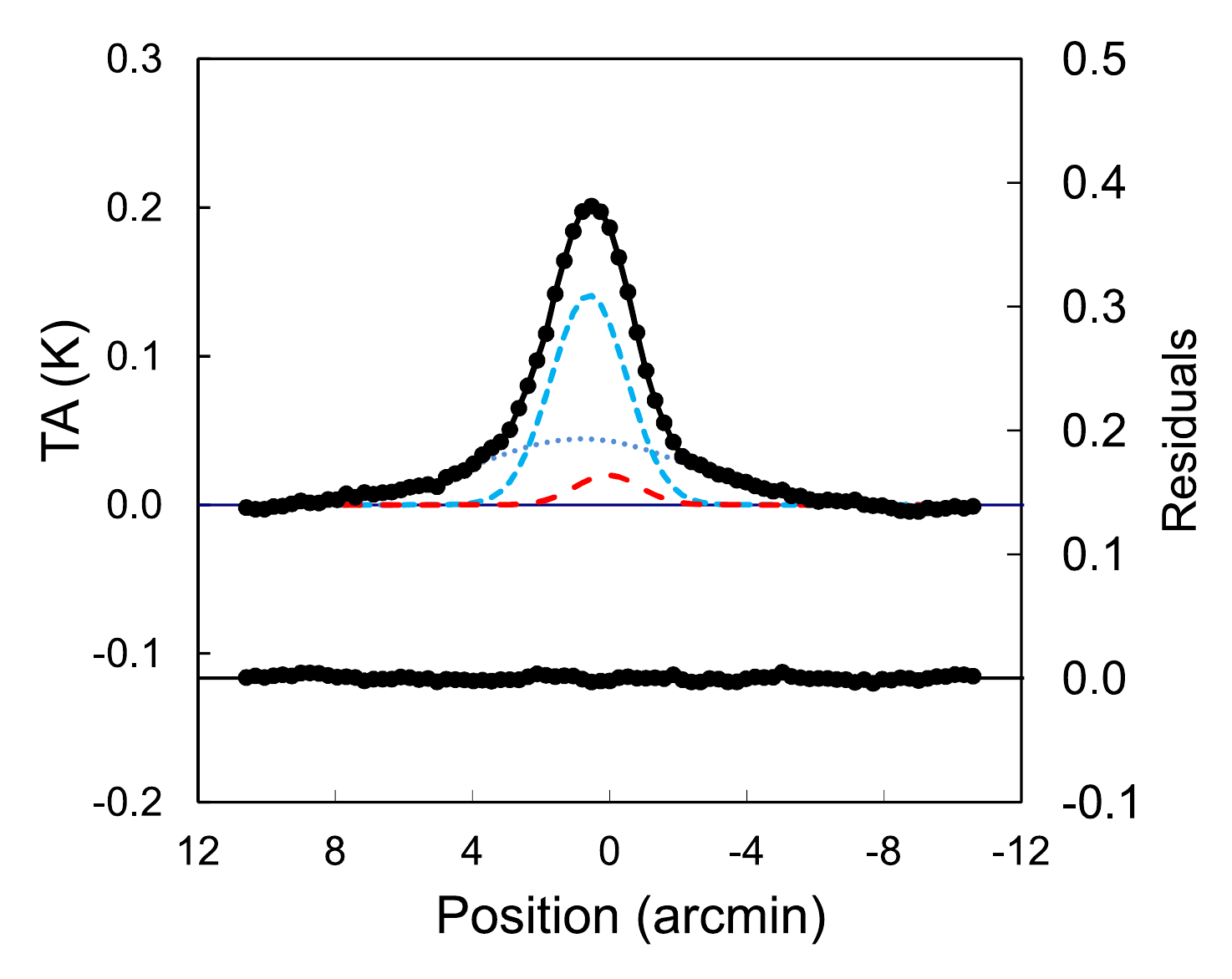}
  \caption{Averaged 7 mm observations (points) during 2007. The pointed and small broken lines represent the gaussian model for the SgrA complex while the large broken line represent the Gaussian contribution of SgrA*. The full line is the sum of the three component model.}
  \label{ADJUST}
\end{figure}

In order to separate the contribution of SgrA* from the surrounding SgrA emission, we first averaged all the observations from the year 2007, and adjusted three gaussians to the data: a point source with HPBW equal to the beam size (2.2 arc min), with two free parameters (peak flux density and central position)  representing  the average flux density of SgrA* during this period, and two other gaussians  with completely free parameters. The result of the best fitting is presented in Table 1; in Fig. \ref{ADJUST}, we show the averaged data (dots), the  adjusted gaussians and the residuals (observations minus model).  The blue and the navy coloured regions in  Fig. \ref{GC} are represented in Fig. \ref{ADJUST} by the two strongest gaussians; the adjusted HPBWs  match very well their angular sizes. The emission of SgrA* is represented by the smaller gaussian. 

During the data reduction process, we also adjusted three gaussians to the observations, with the HPBW of each of them given in Table 1, maintaining constant their relative positions as well as the relative intensities of the two gaussians that represent the contribution of the SgrA complex. The free parameters were the position and intensity of the strongest gaussian, and the intensity of SgrA*.The flux density of SgrA*, calibrated with the SgrA complex, was obtained by dividing its observed flux density by the ratio of the flux densities of the strongest component of the SgrA complex and of its 2007 average value. 

The observations of SgrB2 Main, the quasi-instantaneous calibrator, were made with the same method as those of SgrA*, i.e.  scans in right ascension with 30 arcmin amplitude and 20 s duration.  The observations were made alternately with those of  SgrA* and  with the same integration time.  As for SgrA, two gaussians  were fitted to the averaged observations obtained in 2007, their parameters are presented in the last two rows of Table \ref{Table1}. 

The flux density of SgrA* was also calibrated using the strongest component of SgrB2 Main and he difference of of the results of the two calibrations was taken as  the measurement error.
 %\begin{center}
 \begin{table}
 {\small 
 \caption{Parameters the Gaussian models of SgrA*, SgrA complex and SgrB2.}
 \hfill{} 
 \begin{tabular}{|c|c|c|c|p{\columnwidth}|}
\hline Source & Intensity (K) & HPBW (') & $x_{0}$ (') \\ 
\hline SgrA* & 0.020 & 2.2 & 0.06 \\ 
\hline SgrA (gaussian1) & 0.141 & 2.59 & 0.59 \\ 
\hline SgrA(gaussian2) & 0.044 & 7.30 & 0.80 \\
\hline SgrB2 (gaussian1) & 0.167 & 2.52 & 0.00 \\ 
\hline SgrB2 (gaussian2) & 0.015 & 8.45 & 0.85 \\  
\hline 
\end{tabular}}
\hfill{}
\label{Table1}
\end{table}

%\end{center}

\section{Results and Discussion}
\label{Results}
Although interferometric observations provide flux density measurements with better error bars due to the higher angular resolution, single dish light curves can have better temporal resolution for  variability studies, as  a consequence of more available observation time. This is particularly important for SgrA*, because the previously detected periodicities are of monthly timescale and so, they need long time monitoring to be confirmed. Also, the large day-to-day variability at 22, 43 and 86 GHz, recently reported   by \citet{lu11}, needs a sequence of observations in consecutive days to be detected. This is the first time that SgrA* variability is studied using single dish observations and again, this was only possible because we used the emission of the SgrA complex as a simultaneous calibrator, as discussed in section \ref{Observations}. In Table 2 we show the values of the flux density for the first year of our monitoring. The values for the complete light curve are presented in the online version.

 \begin{table}
 {\small
 \caption{Observations days of SgrA* in 2006}
 \hfill{} 
 \begin{tabular}{|c|c|c|c|p{\columnwidth}|}
\hline Date & JD-2450000 & Flux Density & Error \\
\hline & & (Jy) & (Jy) \\ 
\hline 21-06-2006 & 3907.5 & 1.86 & 0.01 \\ 
\hline 28-06-2006 & 3914.5 & 2.35 & 0.13 \\ 
\hline 29-06-2006 & 3915.5 & 2.09 & 0.04 \\
\hline 30-06-2006 & 3916.5 & 2.10 & 0.12 \\ 
\hline 04-07-2006 & 3921.5 & 1.99 & 0.00 \\
\hline 05-07-2006 & 3922.5 & 2.01 & 0.08 \\
\hline 06-07-2006 & 3923.5 & 2.44 & 0.02 \\
\hline 18-07-2006 & 3935.5 & 1.90 & 0.01 \\
\hline 19-07-2006 & 3936.5 & 2.35 & 0.02 \\
\hline 24-07-2006 & 3941.5 & 1.96 & 0.10 \\
\hline 25-07-2006 & 3942.5 & 2.18 & 0.07 \\
\hline 26-07-2006 & 3943.5 & 2.03 & 0.04 \\ 
\hline 27-07-2006 & 3944.5 & 2.45 & 0.08 \\
\hline 28-07-2006 & 3945.5 & 2.13 & 0.08 \\
\hline 09-08-2006 & 3956.5 & 1.84 & 0.08 \\
\hline 10-08-2006 & 3957.5 & 2.14 & 0.08 \\
\hline 11-08-2006 & 3958.5 & 2.08 & 0.02 \\
\hline 12-08-2006 & 3959.5 & 2.23 & 0.00 \\
\hline 16-08-2006 & 3963.5 & 2.11 & 0.03 \\
\hline 21-08-2006 & 3968.5 & 2.14 & 0.05 \\
\hline 23-08-2006 & 3970.5 & 1.90 & 0.02 \\ 
\hline 
\end{tabular}}
\hfill{}
\label{T2}
\end{table} 

\begin{figure*}
  \includegraphics[width=18 cm]{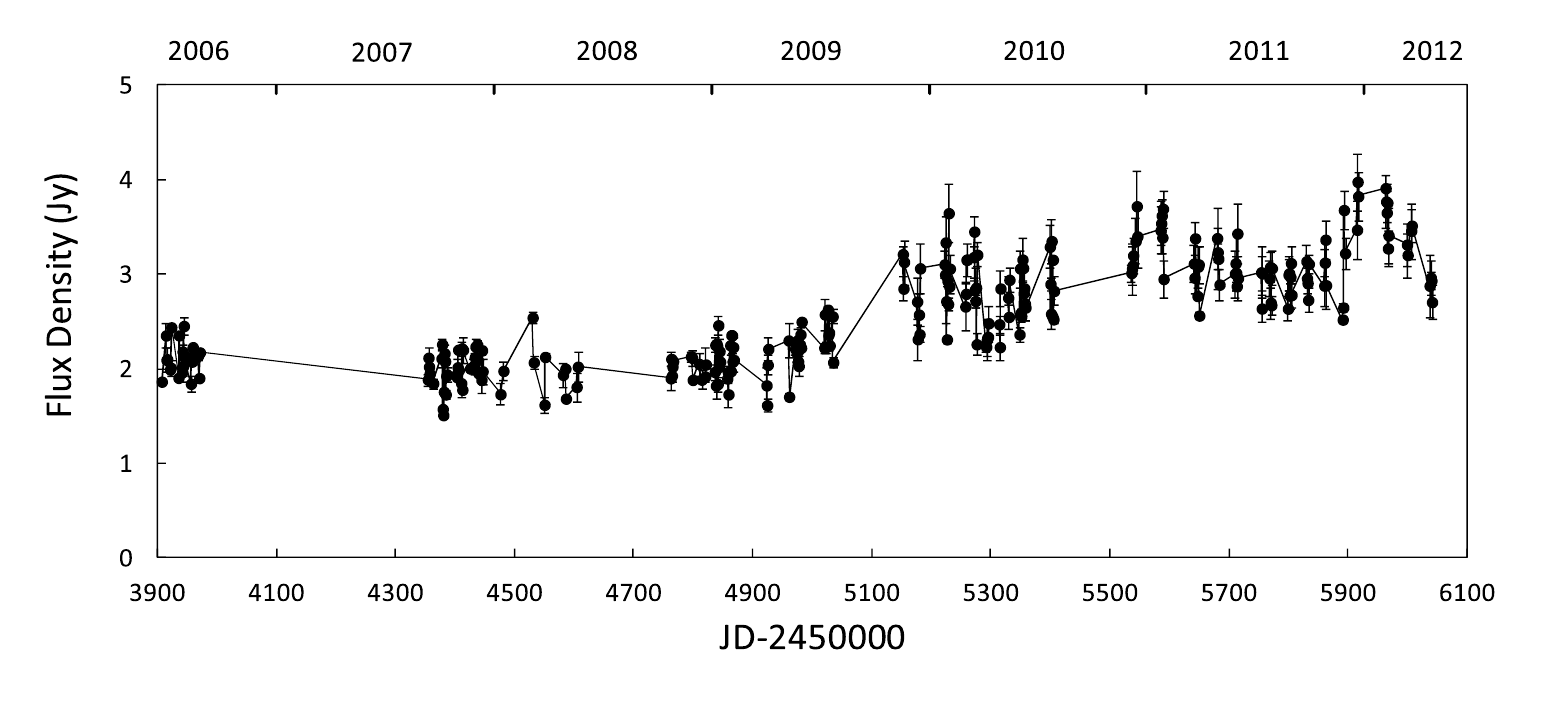}
  \caption{7 mm light curve of SgrA*. The error bars represent the differences obtained using the two calibrators, SgrA complex and SgrB2 main }
  \label{LC}
\end{figure*}  

\subsection{Variability}
\label{variability}
The results of six years monitoring of SgrA* at 7 mm are presented in Fig. \ref{LC}. We detected an increase in the emission after the first semester of 2009; between 2007-2009 the average emission was $2.1 \pm 0.2$ Jy, while it was  $2.8 \pm 0.2$ Jy in 2010 and $3.1 \pm 0.2$ Jy in 2011. At the same time there was an increase in the amplitude of the variability, which had an average value of $1.1$ Jy in 2007-2009 (low phase) and $2.1$ Jy afterwards (high phase). The maximum  flux density in our light curve was 4 Jy, value never reached at 7 mm, although \citet{yus11} detected  3 Jy   with the  VLA  before the beginning of our observations.

\begin{figure}
  \includegraphics[width=\columnwidth]{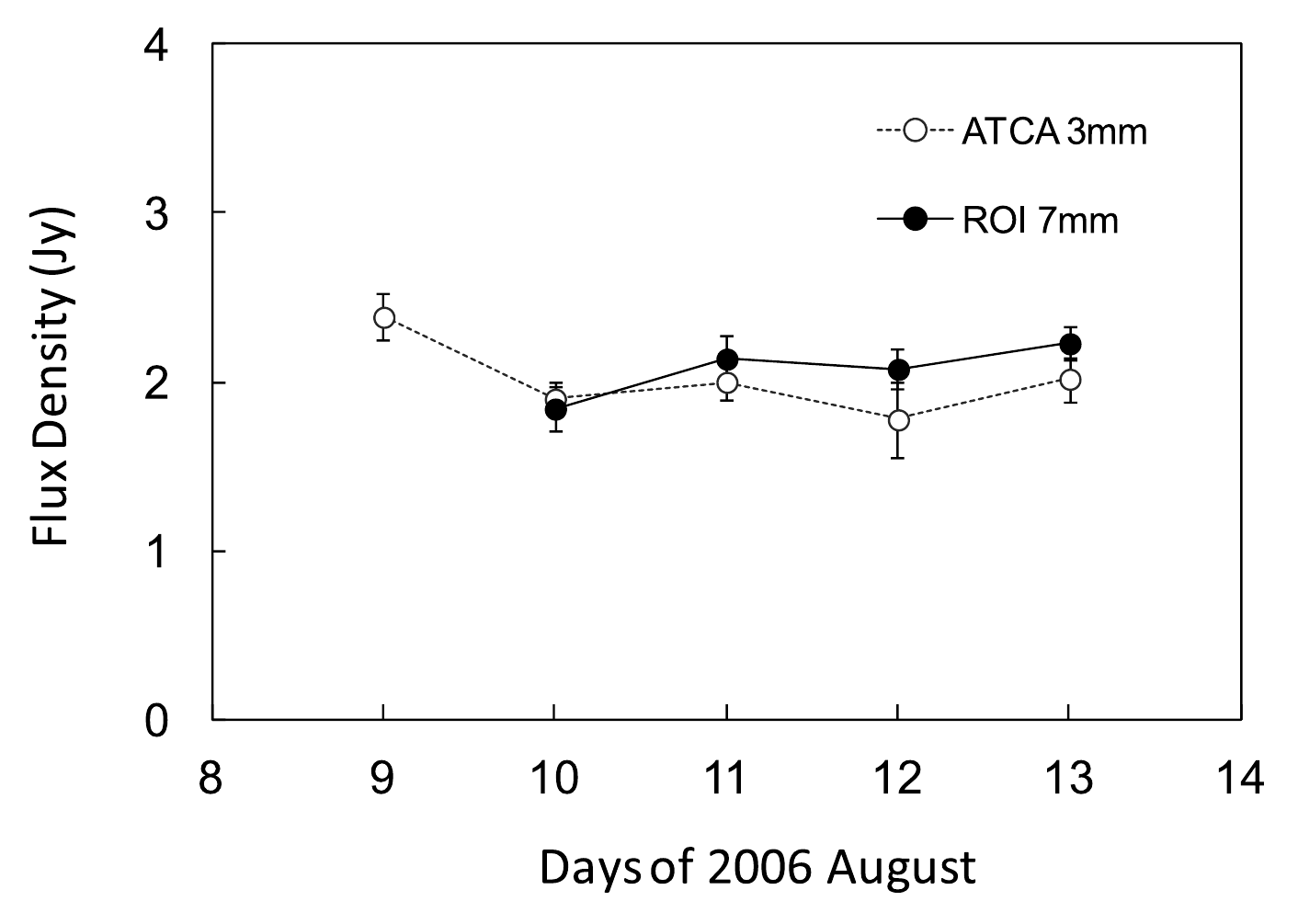}
  \caption{ATCA and Itapetinga simultaneous observations of SgrA*.}
  \label{ROI_ATCA}
\end{figure}

Coincidence in flux density during quasi-simultaneous observations with interferometric techniques, at the same frequency, shows the reliability of our results. In fact, \citet{yus06}, using the VLA for rapid variability studies, reported a 43 GHz flux density  of $1.70 \pm 0.10$ Jy on 2006 July 17, while we found $1.90 \pm 0.01$ Jy on July 18. At other frequencies, \citet{fis11} found a flux density of $2.07 \pm 0.15$ Jy at 1.3 cm on 2009 April 5, while for this day we found  $2.21 \pm 0.13$ Jy at 7 mm. Finally, we have 4 days data coincident with ATCA 3 mm observations \citep{li09}, which show the same variability pattern, as can be seen in Fig. \ref{ROI_ATCA}.  This is the first time that  correlation between variability at these two radio frequencies is detected in SgrA*. The ATCA observations also revealed the existence of IDVs on 2006 August 13. When we looked at our data in a 10 min timescale, we found during the same day a dispersion higher than average; unfortunately, the large error bars of the observations for  short integration times makes it difficult to detected these events without simultaneous interferometric observations.

\label{106}
In our single dish light curve we have 183 observations in consecutive days. We found day-to-day flux density variations larger than 0.5 Jy  in 23 of them, only four times during the low state. In 6 opportunities, the day-to-day variations were even larger than 0.75 Jy.  Similar variability was detected in 2007 at 43 GHz using VLBI observations, when SgrA* emission showed changes in flux density of $0.5$ Jy that coincided with two NIR flares \citep{kun10,lu11}. Unfortunately, there are no infrared data for the epochs at which we detected  strong daily variability. On the other hand, there are 45 consecutive days in which the flux density variation was smaller than our error bars. These events were present both before and after the rise in the light curve, being 16 detected before  2009 December.

\begin{figure}
  \includegraphics[width=\columnwidth]{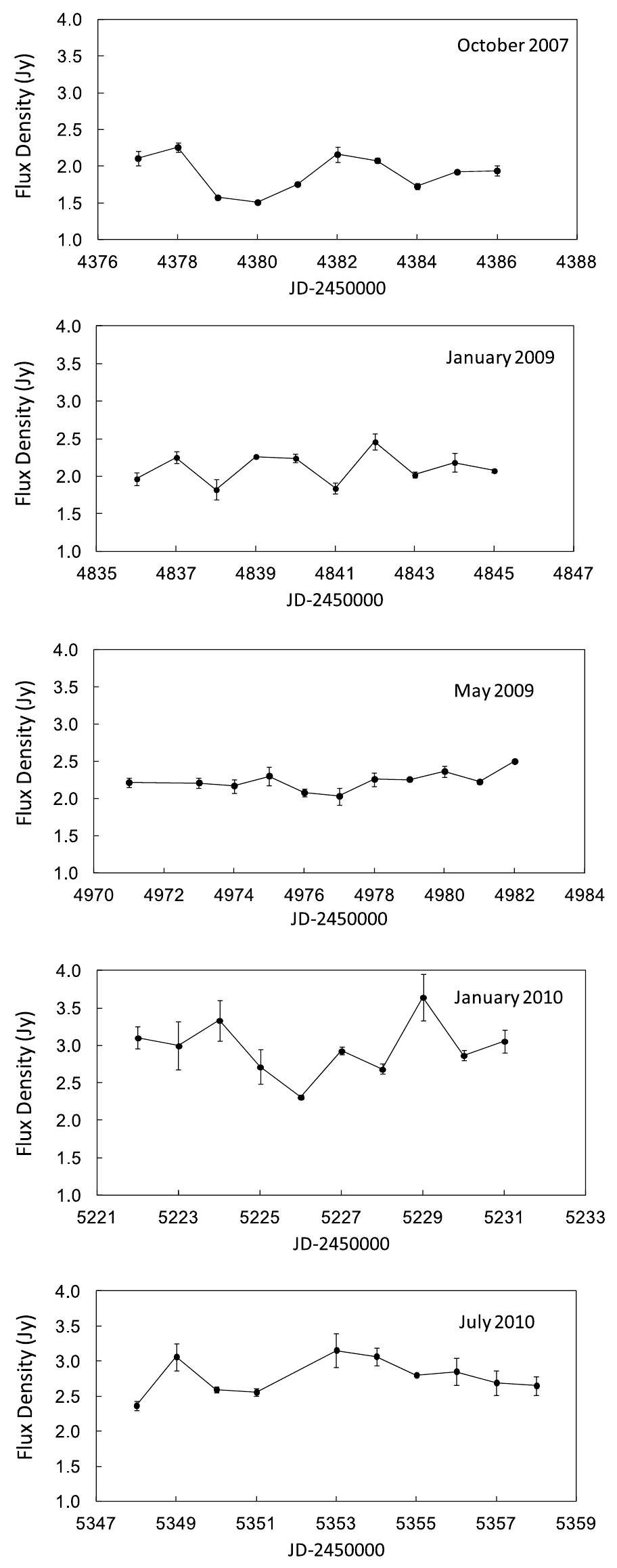}
  \caption{Light curve of SgrA* for five epochs with observations in more than six consecutive days.}
  \label{consecutive}
\end{figure}

In our monitoring program, we have 5 epochs with more than six consecutive days of observation, they are shown in Fig. \ref{consecutive}. Day-to-day variability is present in some of them; the largest occurred in 2010 January, when the flux density changed from a minimum of $1.9 \pm 0.2$ Jy in January 29 to a maximum of $3.7 \pm 0.3$ Jy after 4 days. This happened at the beginning of the high phase in the light curve. Four months later, in  2010 June, the flux density slowly decreased from $3.2 \pm 0.2$ Jy to $2.6 \pm 0.1$ Jy during days 6 and 12. 

The increase in the 7 mm flux density during the last years was also observed at other wavelengths. The RXTE (1.5-12) keV ASM light curve  (Rossi X-Ray Time Explorer, All Sky Monitor\footnote{$http://xte.mit.edu/ASM\_lc.html$}) shows an increase in the SgrA* flux density by a factor of $2.5$  relative to its average value before 2010. SgrA* was never so intense at X-ray since the beginning of the ASM program. In the infrared, the K band intensity also showed an increase that started in 2008, after an intense flare detected in August 5 by the Very Large Telescope \citep{dod11}.

We were able to determine the spectral index $\alpha$ $(S_\nu\propto \nu^\alpha)$ between the 7 mm and H-band  flux densities in two simultaneous observations. In  2006 June 24 and 27, the H-band flux densities were $4 $ and $7$ mJy, respectively, \citep{bre11} while the 7 mm flux densities were  $1.96 \pm 0.01$ and $2.45 \pm 0.05 $ Jy respectively; between those days there was a  monotonous rise in flux at 7mm.  The IR-radio spectral index   was -0.7 for both coincident days, which is compatible with the IR spectral index of $-0.6 \pm 0.2$ found by \citet{hor07}. This coincidence favours the idea that the emission could be produced by the the same non-thermal electron distribution \citep{yua03}.  We  have also 7 mm and K-band simultaneous observations on 2006 June 21, which gave a spectral index of -1.1.

\subsection{Periodicity}

As it was already mentioned, two periods were previously reported in the SgrA* radio emission:  106 days, determined from 20 years of VLA observations at 22 GHz \citep{zha01}, and  57 days found after 600 days of observations with the GBI at 2.3 GHz \citep{fal99}. Our data set is not so long as the VLA database, but has better temporal resolution, while it is longer than the GBI monitoring but with worse temporal resolution. In other words, our light curve is irregularly sampled but without long gaps, which allows a more reliable statistical analysis.

To check the existence of possible periodicities we used the Structure Function Analysis  and the Stellingwerf method. The intensity of the structure function (hereafter SF) is defined as $D(\tau)= \langle [I(t+\tau)-I(t)]^{2} \rangle$. This is a simple method to verify how much the intensity $I$ varies after a time delay $\tau$. If there is no characteristic timescale in the light curve, $D$ will have a random value for each $\tau$, and so there will be no maxima or minima in the structure function. Instead, if there is a typical timescale $\tau^{+}$ in the light curve, $D$ will be large when $\tau = \tau^{+}$ and appear as a maximum in the SF. This method is also useful to find periodicity, because if there is a period $P$, $D$ will have a low value when $\tau$ is equal to an integer number of periods, and this will result in  minima in the SF. 

The Stellingwerf method \citep{ste78} is used to find periodicity in an irregularly sampled series. The test divides the sample in $m$ groups following the phase criterion:
\begin{equation}
\phi_{i}=\frac{t_{i}}{P}-\biggl [\frac{t_{i}}{P} \biggr ]
\end{equation}
where $\phi_{i}$ is the phase, $t_{i}$ is the observation time and $P$ is the guessed period. The brackets correspond to the integer part of the fraction. Each observation day of the sample has a value for the phase  and it is grouped with other days with similar phases.  The variable $\theta$ is defined as the ratio between the sum of the mean square deviation of the fluxes  in each group, and the  mean square deviation of the complete sample. If $P$ did not correspond to the real period,  $\theta$ will be close to 1. If $P$ is indeed the period in the signal, there will be in the same group days with similar flux densities, and so, the sum of the group mean square deviation  will be smaller than that of the complete sample. The number of groups is a free parameter, but higher values of $m$ provide deeper minima in the $\theta$ vs. $P$ curve. However, $m$ can not be too large, because we need a sufficient number of elements in each group to obtain a reliable statistics. We used $m=30$ for all tests in this work.

\begin{figure*}
  \includegraphics[width=14 cm]{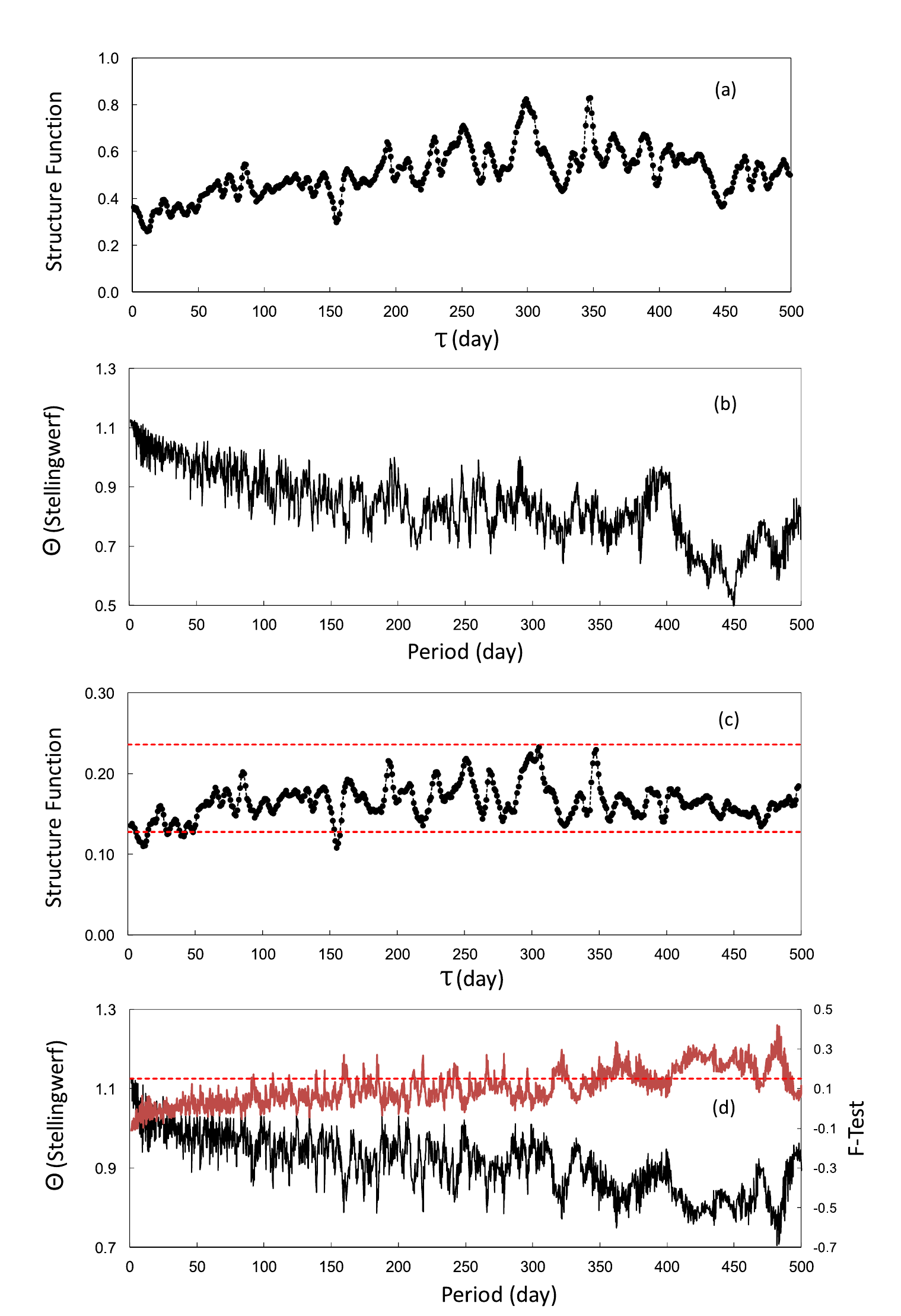}
  \caption{Statistical results for SgrA* 7 mm light curve:  (a) and (b) SF and Stellingwerf results for the original light curve; (c) SF and (d) Stellingwerf and F-test results (in red) after a third-order polynomial was removed from the light curve. The red dashed lines represent the significance limit.}
  \label{Testes}
\end{figure*}

First, we applied both tests to the original light curve; the results are shown in Fig. \ref{Testes} (a) and (b). Then, we applied the tests to the data after removing a third-order polynomial to discount the slow time variation; the results shown in Fig. \ref{Testes} (c) and (d). This last procedure is similar to that used by \citet{zha01}.

\begin{figure*}
  \includegraphics[width=14cm]{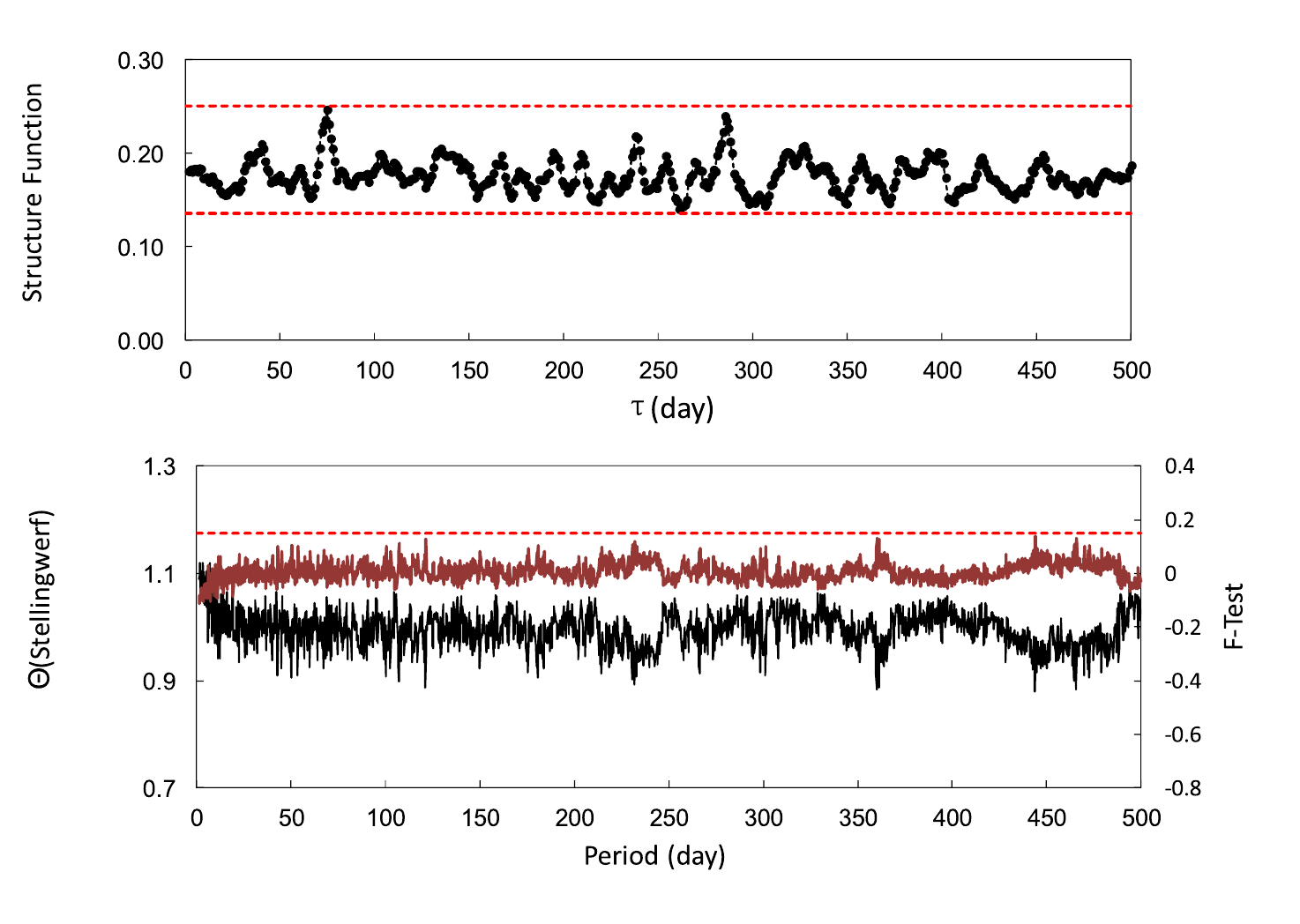}
  \caption{Statistical results for a random signal in the same days as those of our light curve. For the SF, the red dashed lines represent the extremes of the oscillations; for the Stellingwerf method the F-test function is included (in red), the dashed red line is the maximum value of this function}
  \label{random}
\end{figure*}

In both cases we found several maxima and minima. To verify their significance  and check if they are not consequence of sampling problems, we constructed an artificial light curve using random values for the flux density but with the same variability amplitude,  for the days in which SgrA* was observed.   The results of the statistical tests applied to this light curve are presented in Fig. \ref{random}. The range of the oscillations in the SF function lies between $-0.23$ and $+0.43$ of its mean value, which corresponds to $2.5$ and $4.0$ $\sigma$, respectively. We used these values  to limit the significance of the maxima and minima in Fig. \ref{Testes} (c),  shown as red dashed lines. The only minimum deeper than this limit corresponds to $156 \pm 10$ days, but marginal minima are present at the resonant $320 \pm 10$ days period, and at $220 \pm 10$ days, which could be resonant with  the 106 day cycle reported  by \citet{zha01}.

The significance of the Stellingwerf method results can be computed through the F-Test, developed by \citet{kid92}:
\begin{equation}
f=\frac{1-\Theta}{\Theta},
\end{equation}

\noindent
where $f > 0.5$ indicates strong periodicity and $f < 0.25$, the absence or weak periodicity. In fact, the $f$ test applied to the random light curve results shows always $f < 0.15$. This value  is represented as the red dashed line  in Fig. \ref{Testes} (d). We can see  wide 156 and 320 days minima which seem to be significant ($f=0.27$), as well as several other very narrow minima, which we interpret as fluctuations. As discussed in \citet{kid92}, the Stellingwerf method can not reveal periods in a sampling smaller than six times the proposed period. In our light curve, this means periods longer than 350 days.

\begin{figure*}
  \includegraphics[width=15 cm]{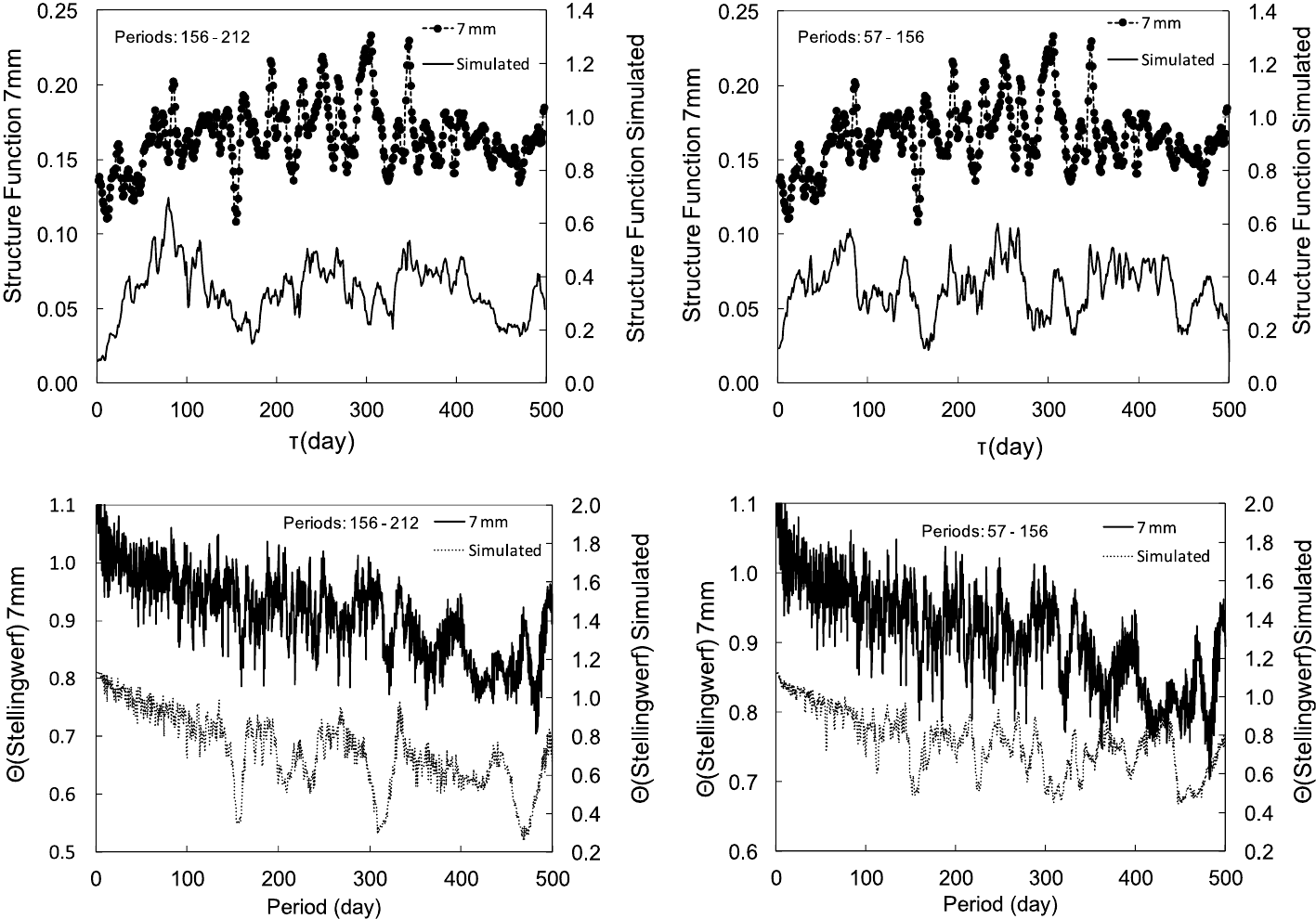}
  \caption{Comparison between the statistic results for the real (upper) and simulated (lower) light curves, for  two different period combinations: 156 and 212 days (left) and 57-156 days (right). Top: Structure Function intensity, bottom: Stellingwerf method.}
  \label{ALC}
\end{figure*}

We conclude from the Structure Function and the Stelingwerf method that are two possible periods: $156 \pm 10$ and $220 \pm 10$ days that do not coincide with any of those  previously  reported: 57 days  \citep{fal99} at 2.3 GHz and and 106 days \citep{zha01} at 22 GHz. However, we speculated if the existence of two beating periods could be the reason for the different results obtained by different authors. To check this hypothesis, we constructed  artificial light curves as combination of two sine square functions, leaving their relative amplitudes as a free parameter and adding a random signal with  amplitude similar to that of the periodic functions. We tried all possible combinations of 57, 106, 156, 212, 220 periods. We  used 212 days because it is exactly a resonance of  the 106 day cycle. Two combinations reproduced  our statistical results: 156 and 212 days, and the 57 and 156 days;  the amplitudes of the 156 days periodic sine square functions were 1.2 and 1.5 times larger than that of the 57  and  212 days periodic signals, respectively. The statistical tests are compared in Fig. \ref{ALC}. The 156 days minimum was not found by any combination of other periods, and the result of using 220 or 212 days period are very similar.

\begin{figure}
  \includegraphics[width=\columnwidth]{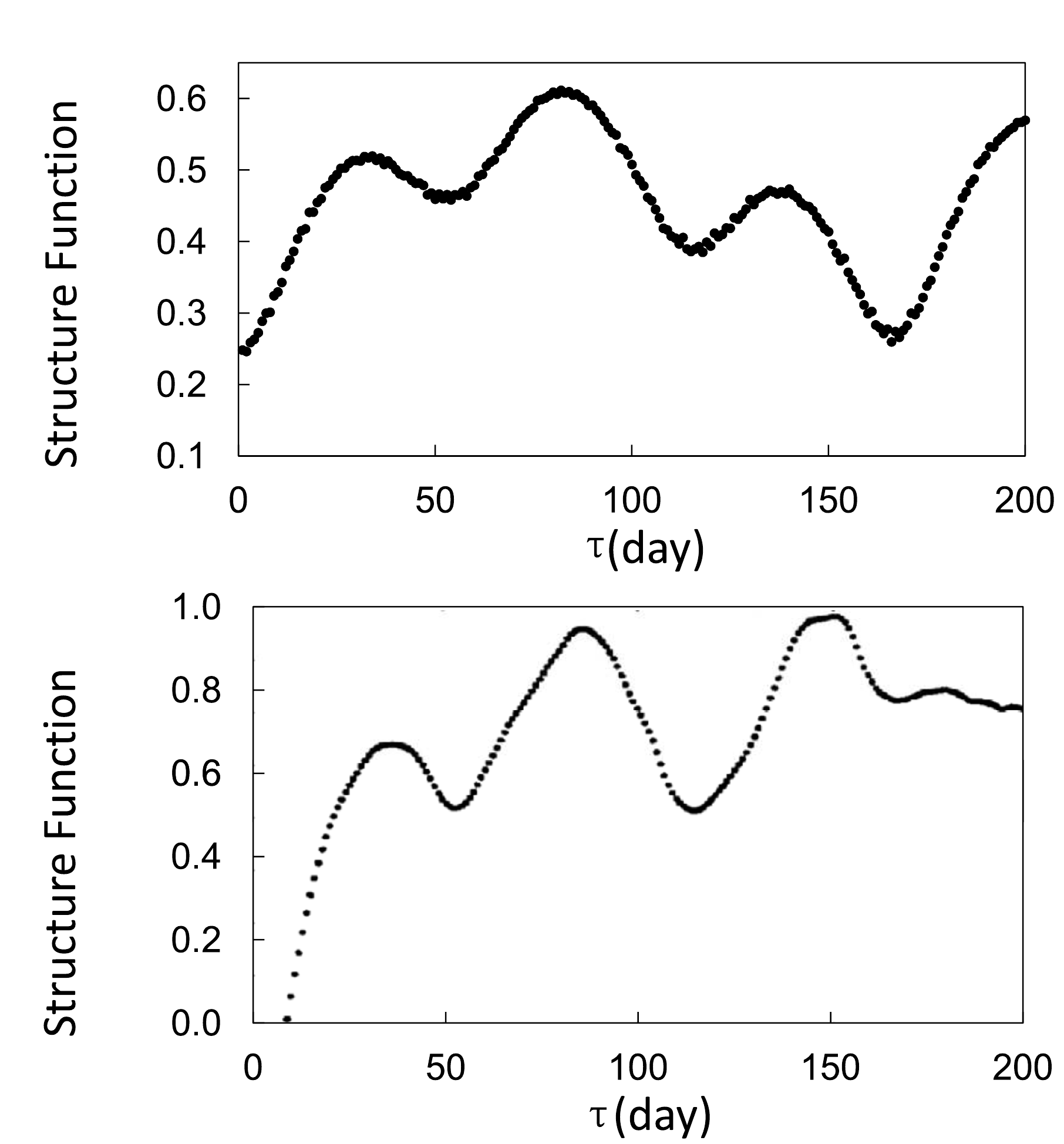}
  \caption{Top: SF results for the simulated two period light curve (57 and 156 days) with daily time resolution. Bottom: the SF results obtained by \citet{fal99} for the GBI monitoring at 2.3 GHz.  The SF of the \citet{fal99} data are normalized by the highest peak. }
  \label{F99}
\end{figure}

The minimum that would correspond to the 57 days period did not appear in any of the combinations, even if this period was used in the input data. We believe that this could be a consequence of not having enough temporal  resolution added to the existence of large amplitude short timescale fluctuations. Therefore, we repeated our calculations using our simulated light curve for the 57 and 156 days periods with  daily data points, and in fact, the presence of the 57 days minimum can be seen in Fig. \ref{F99}. It is important to notice that the  SF of the simulated light curve is in very good agreement with the SF found by \citet{fal99} for the 2.3 GHz data and that the larger amplitude of the second peak seen in both SFs is a consequence of the existence of these two periodicities. This was not the case when the combination of 156 and 212 days periodic signals were used, even with  daily resolution.

In other words, only the combination of 57 and 156 days reproduce both our statistical tests and Falcke's. To verify the range of possible periods around  57 days that can still reproduce  Fig. \ref{ALC}, we repeated the tests combining the 156 day period with different periods between 51 and 61 days. If the second period is smaller than 55, there is no minimum in the SF around 220 days. Otherwise, if the second period is larger than 59 days, the minimum in 156 days is not deep enough. This indicates that a periodic signal of of $57 \pm 2$ days mixed with other of 156 days can reproduce the $156 \pm 10$ and $220 \pm 10$ minima revealed by our statistical tests. 

Considering the existence of the $57 \pm 2$ period, we can interpret the 106 day cycle reported by \citet{zha01} as a resonance of that period. To check this hypothesis we modulated our light curve with a 106 day period, and compared the result with that of \citet{zha01}. We  found  a similar pattern for both cycles, but without the two large peaks on days 75 and 100 present in the 13 mm cycle. The peaks can be due to the low number of observation days  in the VLA light curve. 
    
Finally, we must mention that  no periodicity was reported by \citet{mac06} for the VLA data obtained by \cite{her04}. To understand that we  calculated the SF for these data and we did not find any periodicity either. Again, the reason is that the VLA light curve has not enough  time coverage  to detect monthly periodicities.

\subsection{The infalling ionized gas cloud}

\begin{figure}
  \includegraphics[width=\columnwidth]{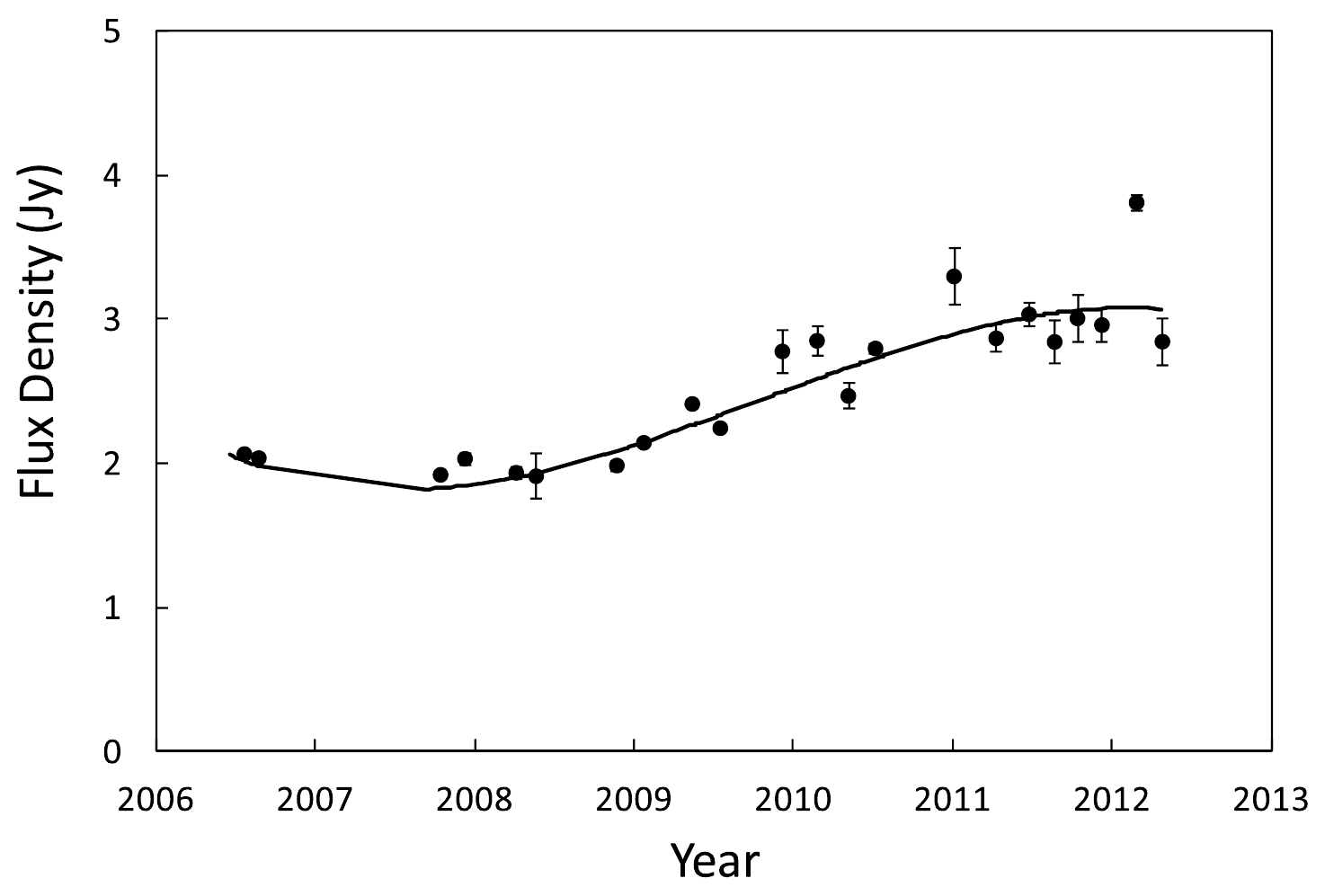}
  \caption{7mm light curve with 60 days binning and the third order polynomial removed from the light curve (continue line)}
  \label{BIN60}
\end{figure}

As presented in subsection 3.1, variability of SgrA* was detected at 7 mm in timescales of days and years. The latter consisted of a continuous increase in flux density that started in 2008, as can be seen in Fig.\ref{BIN60}, where the radio data are averaged in 60 days bins.  Coincidently,  \citet{gil12} detected a dense gas cloud that is moving in the direction of SgrA* in a highly eccentric orbit and will reach its minimum approach in 2013.5. The almost spherical cloud of 15 mas $(2\times 10^{15}$ cm) observed in Br$\gamma$ in 2008, is being tidally disrupted and in 2011 presented a  tail  elongated in the orbital direction of about 200 mas.

To investigate if the  increase in the 7 mm flux density between 2008 and 2011 could be the result of the appearance of the disrupted tail, we calculated the flux density of an ionized optically thick cloud of temperature $T$  that subtends a solid angle $\Omega=\pi\theta^2$:

\begin{equation}
S_{\lambda}=\frac{2kT}{\lambda ^2}\Omega
\label{S}
\end{equation}

\noindent where $S_\lambda$ is the flux density at wavelength $\lambda$, and $k$ is the Boltzmann constant.

Assuming $T=10^4$ K and $\theta=150$ mas, the size of the observed tail, we obtain $S_{7 \rm mm} = 1$ Jy, in agreement with our observations. We could assume then that the increase in the observed 7 mm flux density between 2008 and 2011 corresponded to an increase in the solid angle $\Omega$. This increase seemed to stop around 2011, which could be interpreted as the cloud becoming optically thin at that epoch. Assuming the optical depth $\tau_{7 \rm mm}=1$, we can calculate the emission measure $EM=n_{H}^{2}L$ of the cloud from:

\begin{equation}
\tau_{\lambda}=2 \times 10^{-23}n_{H}^{2}L\frac{\lambda^2}{T ^{3/2}}g_{ff}
\end{equation}

\noindent
For $T=10^4$ K, the  Gaunt factor is $g_{ff}\sim 7$, and $EM=1.4\times 10^{28}$ cm$^{-5}$.

If we assume a density of $10^7$ cm$^{-3}$ for the cloud, as obtained from numerical simulations of the evolution of a disrupted cloud \citep{bur12}, we obtained a length $L=1.4\times 10^{14}$ along the line of sight, showing that the  tail in the orbital plane could be formed by a thin layer of gas.

If the 7 mm radio emission is due to free-free emission, we should expect a corresponding flux density in the IR, given by

\begin{equation}
S_{\lambda}=5.4\times10^{-16}T^{-0.5}g_{ff}(\lambda, T)\Omega  EM \exp(\frac{-hc}{\lambda kT}) \,\,\,\, {\rm Jy.}
\end{equation}

For the IR, $g_{ff}(\lambda, T)\sim 1$; resulting for the $Ks$ band a flux density of $S_{Ks}\sim 0.057$ Jy. Although this value is much higher than any flux density detected in the SgrA* region, it must be noticed that it corresponds to an extended emission, therefore the signal should be  diluted in an area of $150\times 150$ mas$^2$, which represents the size of the tail. Considering a pixel size of 13 or 27 mas per pixel, typical of the VLT data \citep{dod11,gil12} we obtain 0.43 and 1.8 mJy per pixel, which could have been overlooked in the data.

However, the increase in the amplitude of the fluctuations in timescales of days simultaneously with the increase in the total flux density that started around 2009, favours the hypothesis that is intrinsic to SgrA*. In fact,this relation between rms fluctuations and flux density was already detected in active galaxies and X-ray binaries at X-rays \citep{utt01}and optical wavelengths \citep{gan09} when they are in a luminous phase, which lead \citet{hei12} to claim that this can be an overall propriety of the accretion flow in compact sources. In the case of SgrA*, the increase in the accretion rate could be related to the influence of the infalling cloud or to the winds of faint stars during their periastron passage \citep{lob04}, as those reported by \citet{dod11}.

\section{Conclusion}
\label{conclusion}

In this paper we presented the results of single dish  monitoring of SgrA* at 7 mm  during a six year period. 
Great care was taken to separate the contribution of SgrA* from the surrounding emission, the latter used, together with the HII region  SgrB2, as an instantaneous calibrator.
The reliability of our results was verified by comparing them with interferometric observations at coincident epochs.

\begin{figure}
  \includegraphics[width=\columnwidth]{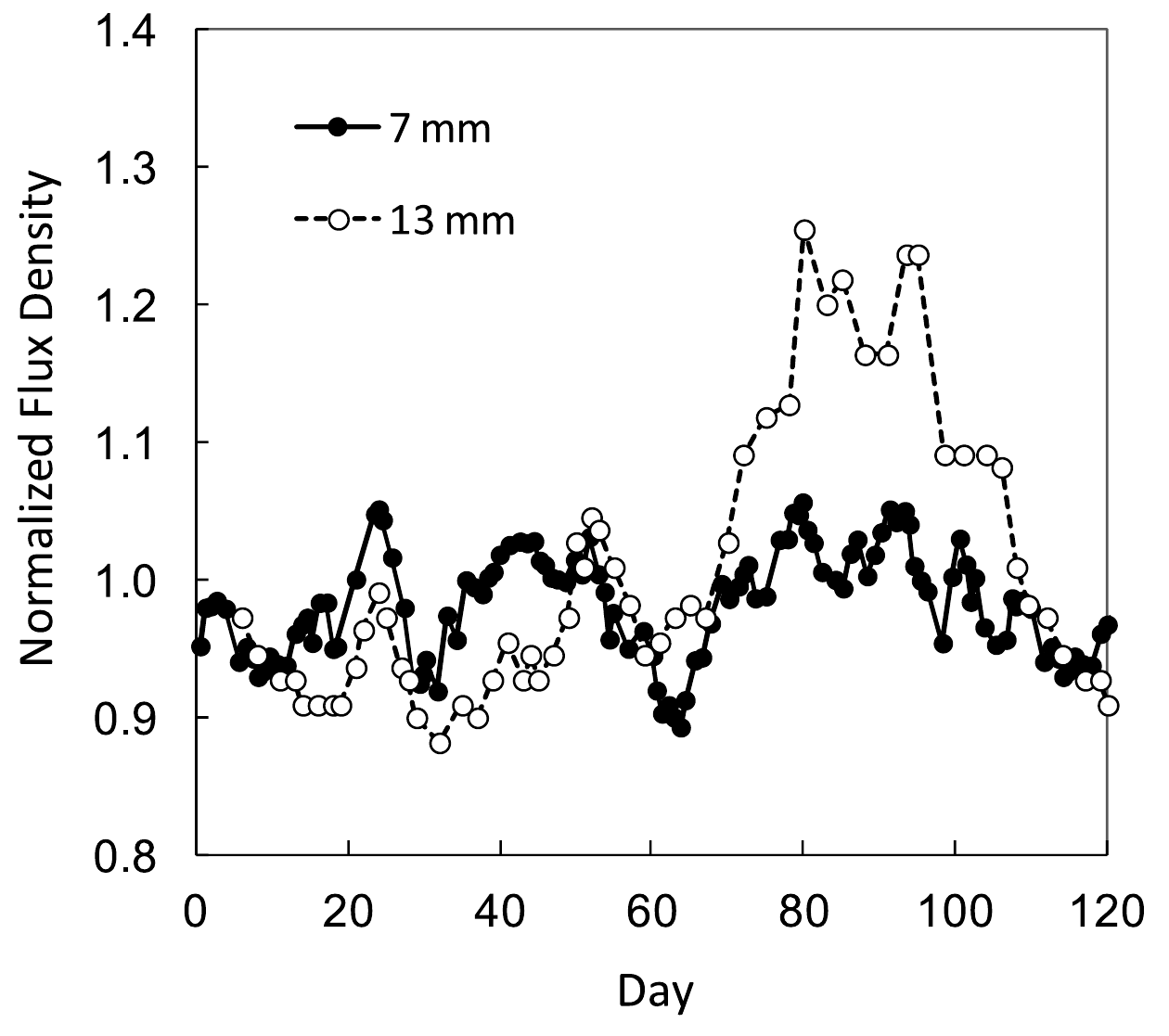}
  \caption{106 day modulation of the 7 and the 13 mm light curves}
  \label{7_13}
\end{figure} 

During the observing campaign, several epochs with consecutive day observations  allowed the study of variability in daily as well as longer timescales. The resulting light curve shows a slow variation in a timescale of years, a random component with shorter timescale and a day-to-day variability with large amplitude variations between two consecutive days. 

We summarize the variability result as follow:

\begin{flushleft}
{
\begin{enumerate}
\item SgrA* emission at 7mm has been growing up through the years since 2008, this behaviour coincides with an  increase in X-ray emission, and there is also some evidence of  high activity in the  infrared stating at the beginning of 2009. 
\item SgrA* light curve presented episodes of high amplitude day-to-day variability that were more common in the high phase;
\item The variability at 7 mm showed the same pattern than that at 3 mm  during four coincident days in 2006;
\item Simultaneous 7 mm and IR data revels that the radio-IR spectral index is similar to  the IR spectral index, suggesting that the emission at that epoch was probably produced by the same electron population; 
\end{enumerate}
}
\end{flushleft}

We investigate the possibility that the increase at 7 mm flux density that started in 2008 could be related to the appearance of a Br$\gamma$ emitting gaseous  tail of tidally disrupted material, arising from the compact gas cloud that is falling towards  SgrA* in a highly eccentric orbit \citep{gil12}. Even though it is possible to explain the 7 mm flux density increase as free-free emission from such a cloud, the predicted IR emission was not observed. However, the increase of the fluctuations together with the flux density at 7 mm , favours the intrinsic origin of this extra emission and is compatible with the rms/flux relation found in other compact sources \citep{utt01}. 

In search for periodicity, the light curve was statistically analysed using the structure function and the Stellingwerf method. 
The result can be summarized as follows:

\begin{flushleft}
{
\begin{enumerate}
\item The statistical tests realized in this work revealed the existence of two minima: $156 \pm 10$ and $220 \pm 10$ days, but did not show any minima related to the 57 or 106 day periods reported by \citet{fal99} and \citet{zha01}.To test the reliability of our results we constructed an artificial light curve with random signals in the same observation days; no periodicity was found, showing that the existence of periodicity was not due to sampling effects.
\item We investigate if the divergence between the detected periods can be a consequence of the existence of two periodic signals. For that we simulated a light curve,  as a combination of two quadratic sinusoidal functions, with data at the same observation days as those of our light curve, and verified which periods and relative amplitudes better reproduced our statistical results. We concluded that two combinations of periods can explain our statistical results: 156 plus 212 days, and 57 plus 156 days. The amplitudes of the 156 days periodic functions were 1.2 and 1.5 times larger than that of the 57 and 212 days periodic signals, respectively.
\item To test the influence of sampling in the detected periodicity, we constructed a simulated light curve for the 57 and 156 days periods with daily data points, using the same relative amplitudes as in the previous best test. The result of the structure function was in very good agreement with that found by \citet{fal99} for  2.3 GHz, which revealed the 57 day periodicity. 
\item We also modulated our light curve with a 106 day period, to compare it with \citet{zha01} results at 13 mm.  The modulated light curves shows a similar pattern but without the two large peaks on days 75 and 100 of the cycle. This similarity can be explained as a consequence of a resonance of the 57 day period.

\end{enumerate}
}
\end{flushleft}

We conclude that two periodic signals added to a random component can reproduce statistical results similar to what it was obtained for the SgrA* emission. The combination of $156 \pm 10$ and $57 \pm 2$ day periodicity is what better explains our results. If real, these two periods may be explained by precession and/or nutation of a accretion disk. The precessing disk in SgrA* could be consequence of the the Bardeen-Peterson effect \citep{bar75}, as discussed by \citet{liu02}. In fact, this process can lead to a hundred day period in the SgrA* radio emission \citep{cap04,roc05}.

\section*{Acknowledgments}

This work was partially supported by the Brazilian research agencies FAPESP and CNPq. We thank Paulo Penteado and Graziela Keller Rodrigues for useful comments about computational and statistics methods.

%\bibliographystyle{pp_mnras}
%\bibliography{clumping3}

\begin{thebibliography}{99}
\bibitem[Abraham \& Kokubun(1992)]{abr92} Abraham, Z., \& Kokubun, F.\ 1992, \aap, 257, 831
\bibitem[Baganoff et al.(2001)]{bag01} Baganoff, F.~K., Bautz, M.~W., Brandt, W.~N., et al.\ 2001, \nat, 413, 45 
\bibitem[Baganoff et al.(2003)]{bag03} Baganoff, F.~K., Maeda, Y., Morris, M., et al.\ 2003, \apj, 591, 891
\bibitem[Balick \& Brown(1974)]{bal74} Balick, B., \& Brown, R.~L.\ 1974, \apj, 194, 265
%\bibitem[Bondi(1952)]{bon52} Bondi, H.\ 1952, \mnras, 112, 195
\bibitem[Bremer et al.(2011)]{bre11} Bremer, M., Witzel, G., Eckart, A., et al.\ 2011, \aap, 532, A26
\bibitem[Bardeen \& Petterson(1975)]{bar75} Bardeen, J.~M., \& Petterson, J.~A.\ 1975, \apjl, 195, L65
\bibitem[Broderick \& Loeb(2005)]{bro05} Broderick, A.~E., \& Loeb, A.\ 2005, \mnras, 363, 353
\bibitem[Broderick \& Loeb(2006)]{bro06} Broderick, A.~E., \& Loeb, A.\ 2006, \mnras, 367, 905
\bibitem[Burkert et al.(2012)]{bur12} Burkert, A., Schartmann, M., Alig, C., et al.\ 2012, \apj, 750, 58
\bibitem[Caproni et al.(2004)]{cap04} Caproni, A., Mosquera Cuesta, H.~J., \& Abraham, Z.\ 2004, \apjl, 616, L99 
\bibitem[Coker et al.(1999)]{cok99} Coker, R., Melia, F., \& Falcke, H.\ 1999, \apj, 523, 642
\bibitem[Cuadra et al.(2006)]{cua06} Cuadra, J., Nayakshin, S., Springel, V., \& Di Matteo, T.\ 2006, Revista Mexicana de Astronomia y Astrofisica Conference Series, 26, 139
\bibitem[Do et al.(2009)]{do09} Do, T., Ghez, A.~M., Morris, M.~R., et al.\ 2009, \apj, 691, 1021
\bibitem[Dodds-Eden et al.(2011)]{dod11} Dodds-Eden, K., Gillessen, S., Fritz, T.~K., et al.\ 2011, \apj, 728, 37   
\bibitem[Doeleman et al.(2008)]{doe08} Doeleman, S.~S., Weintroub, J., Rogers, A.~E.~E., et al.\ 2008, \nat, 455, 78
%\bibitem[Doeleman et al.(2009)]{doe09} Doeleman, S.~S., Fish, V.~L., Broderick, A.~E., Loeb, A., \& Rogers, A.~E.~E.\ 2009, \apj, 695, 59
\bibitem[Duschl \& Lesch(1994)]{dus94} Duschl, W.~J., \& Lesch, H.\ 1994, \aap, 286, 431  
\bibitem[Eckart \& Genzel(1996)]{eck96} Eckart, A., \& Genzel, R.\ 1996, \nat, 383, 415
\bibitem[Edelson \& Krolik(1988)]{ede88} Edelson, R.~A., \& Krolik, J.~H.\ 1988, \apj, 333, 646 
\bibitem[Eisenhauer et al.(2003)]{eis03} Eisenhauer, F., Sch{\"o}del, R., Genzel, R., et al.\ 2003, \apjl, 597, L121
\bibitem[Falcke(1999)]{fal99} Falcke, H.\ 1999, The Central Parsecs of the Galaxy, 186, 113
\bibitem[Falcke \& Markoff(2000)]{fal00} Falcke, H., \& Markoff, S.\ 2000, \aap, 362, 113
\bibitem[Falcke et al.(2000)]{fal00b} Falcke, H., Melia, F., \& Agol, E.\ 2000, \apjl, 528, L13 
\bibitem[Fish et al.(2011)]{fis11} Fish, V.~L., Doeleman, S.~S., Beaudoin, C., et al.\ 2011, \apjl, 727, L36
\bibitem[Gandhi(2009)]{gan09} Gandhi, P.\ 2009, \apjl, 697, L167  
\bibitem[Genzel et al.(2003)]{gen03} Genzel, R., Sch{\"o}del, R., Ott, T., et al.\ 2003, \nat, 425, 934
\bibitem[Ghez et al.(2004)]{ghe04} Ghez, A.~M., Wright, S.~A., Matthews, K., et al.\ 2004, \apjl, 601, L159 
\bibitem[Gillessen et al.(2009)]{gil09} Gillessen, S., Eisenhauer, F., Trippe, S., et al.\ 2009, \apj, 692, 1075
\bibitem[Gillessen et al.(2012)]{gil12} Gillessen, S., Genzel, R., Fritz, T.~K., et al.\ 2012, \nat, 481, 51
\bibitem[Heil et al.(2012)]{hei12} Heil, L.~M., Vaughan, S., \& Uttley, P.\ 2012, \mnras, 422, 2620  
\bibitem[Herrnstein et al.(2004)]{her04} Herrnstein, R.~M., Zhao, J.-H., Bower, G.~C., \& Goss, W.~M.\ 2004, \aj, 127, 3399
\bibitem[Hornstein et al.(2007)]{hor07} Hornstein, S.~D., Matthews, K., Ghez, A.~M., et al.\ 2007, \apj, 667, 900 
\bibitem[Jolley \& Kuncic(2008)]{jol08} Jolley, E.~J.~D., \& Kuncic, Z.\ 2008, \apj, 676, 351
\bibitem[Kidger et al.(1992)]{kid92} Kidger, M., Takalo, L., \& Sillanpaa, A.\ 1992, \aap, 264, 32 
\bibitem[Kuncic \& Bicknell(2004)]{kun04} Kuncic, Z., \& Bicknell, G.~V.\ 2004, \apj, 616, 669
\bibitem[Kuncic \& Bicknell(2007)]{kun07} Kuncic, Z., \& Bicknell, G.~V.\ 2007, \apss, 311, 127
\bibitem[Kunneriath et al.(2010)]{kun10} Kunneriath, D., Witzel, G., Eckart, A., et al.\ 2010, \aap, 517, A46 
\bibitem[Li et al.(2009)]{li09} Li, J., Shen, Z.-Q., Miyazaki, A., et al.\ 2009, \apj, 700, 417
\bibitem[Liu \& Melia(2002)]{liu02} Liu, S., \& Melia, F.\ 2002, \apjl, 573, L23
\bibitem[Loeb(2004)]{lob04} Loeb, A.\ 2004, \mnras, 350, 725 
\bibitem[Lu et al.(2008)]{lu08} Lu, R.-S., Krichbaum, T.~P., Eckart, A., et al.\ 2008, Journal of Physics Conference Series, 131, 012059
\bibitem[Lu et al.(2011)]{lu11} Lu, R.-S., Krichbaum, T.~P., Eckart, A., et al.\ 2011, \aap, 525, A76 
\bibitem[Macquart \& Bower(2006)]{mac06} Macquart, J.-P., \& Bower, G.~C.\ 2006, \apj, 641, 302
%\bibitem[Mapelli et al.(2012)]{map12} Mapelli, M., Hayfield, T., Mayer, L., \& Wadsley, J.\ 2012, \apj, 749, 168 
\bibitem[Markoff et al.(2007)]{mar07} Markoff, S., Bower, G.~C., \& Falcke, H.\ 2007, \mnras, 379, 1519
%\bibitem[Marrone et al.(2008)]{mar08} Marrone, D.~P., Baganoff, F.~K., Morris, M.~R., et al.\ 2008, \apj, 682, 373
\bibitem[Melia(1992)]{mel92} Melia, F.\ 1992, \apjl, 387, L25 
\bibitem[Melia(1994)]{mel94} Melia, F.\ 1994, \apj, 426, 577   
\bibitem[Melia \& Falcke(2001)]{mel01} Melia, F., \& Falcke, H.\ 2001, \araa, 39, 309
\bibitem[Melo et al.(2009)]{mel09} Melo, I., Tom{\'a}{\v s}ik, B., Torrieri, G., et al.\ 2009, \prc, 80, 024904
\bibitem[Menten et al.(1997)]{men97} Menten, K.~M., Reid, M.~J., Eckart, A., \& Genzel, R.\ 1997, \apjl, 475, L111 
\bibitem[Meyer et al.(2008)]{mey08} Meyer, L., Do, T., Ghez, A., et al.\ 2008, \apjl, 688, L17
\bibitem[Mo{\'s}cibrodzka et al.(2006)]{mos06} Mo{\'s}cibrodzka, M., Das, T.~K., \& Czerny, B.\ 2006, \mnras, 370, 219
\bibitem[Narayan et al.(1998)]{nar98} Narayan, R., Mahadevan, R., Grindlay, J.~E., Popham, R.~G., \& Gammie, C.\ 1998, \apj, 492, 554
\bibitem[Prescher \& Melia(2005)]{pre05} Prescher, M., \& Melia, F.\ 2005, \apj, 632, 1048
\bibitem[Rockefeller et al.(2005)]{roc05} Rockefeller, G., Fryer, C.~L., \& Melia, F.\ 2005, \apj, 635, 336
\bibitem[Sch{\"o}del et al.(2002)]{sch02} Sch{\"o}del, R., Ott, T., Genzel, R., et al.\ 2002, \nat, 419, 694
\bibitem[Stellingwerf(1978)]{ste78} Stellingwerf, R.~F.\ 1978, \apj, 224, 953
\bibitem[Uttley \& McHardy(2001)]{utt01} Uttley, P., \& McHardy, I.~M.\ 2001, \mnras, 323, L26  
\bibitem[van der Laan(1966)]{van66} van der Laan, H.\ 1966, \nat, 211, 1131
\bibitem[Yuan et al.(2003)]{yua03} Yuan, F., Quataert, E., \& Narayan, R.\ 2003, \apj, 598, 301
\bibitem[Yuan et al.(2002)]{yua02} Yuan, F., Markoff, S., \& Falcke, H.\ 2002, \aap, 383, 854
\bibitem[Yusef-Zadeh et al.(2000)]{yus00} Yusef-Zadeh, F., Melia, F., \& Wardle, M.\ 2000, Science, 287, 85 
\bibitem[Yusef-Zadeh et al.(2006)]{yus06} Yusef-Zadeh, F., Roberts, D., Wardle, M., Heinke, C.~O., \& Bower, G.~C.\ 2006, \apj, 650, 189
\bibitem[Yusef-Zadeh et al.(2008)]{yus08} Yusef-Zadeh, F., Wardle, M., Heinke, C., et al.\ 2008, \apj, 682, 361
\bibitem[Yusef-Zadeh et al.(2011)]{yus11} Yusef-Zadeh, F., Wardle, M., Miller -Jones, J.~C.~A., et al.\ 2011, \apj, 729, 44
\bibitem[Zhao et al.(2001)]{zha01} Zhao, J.-H., Bower, G.~C., \& Goss, W.~M.\ 2001, \apjl, 547, L29
\bibitem[Zhao et al.(1992)]{zha92} Zhao, J.-H., Goss, W.~M., Lo, K.-Y., \& Ekers, R.~D.\ 1992, Relationships Between Active Galactic Nuclei and Starburst Galaxies, 31, 295  
 
 
\end{thebibliography}
\end{document}